\documentclass{article}

\def\xxinput#1{\input#1}

\xxinput{vsolj02.sty}

\usepackage{longtable}
\usepackage[dvipdfmx]{graphicx}
\usepackage[comma,colon]{natbib}

\usepackage[OT2,T1]{fontenc}

\def\cite{\citealt}
\setcitestyle{aysep={}}

\def\commenta{$^*$}
\def\commentb{$^\dagger$}
\def\commentc{$^\ddagger$}

\newcounter{author}
\def\altaffilmark#1{$^{#1}$}
\def\altaffiltext#1{$^{#1}$\,}

\def\authorcount#1#2{{\refstepcounter{author}\label{#1}
                     \altaffiltext{\ref{#1}}{#2}}}

\begin{document}

\begin{center}

\title{LS And: WZ Sge-type outburst first time since the 1971 discovery}

\author{
        Taichi~Kato\altaffilmark{\ref{affil:Kyoto}}
}
\email{tkato@kusastro.kyoto-u.ac.jp}

\authorcount{affil:Kyoto}{
     Department of Astronomy, Kyoto University, Sakyo-ku,
     Kyoto 606-8502, Japan}

\end{center}

\begin{abstract}
\xxinput{abst.inc}
\end{abstract}

   LS And was discovered by \citet{vandenber73lsand} in
the region of M~31 (named ``m'' in their paper).
\citet{vandenber73lsand} stated that the object was visible
only on a blue and on a yellow plate taken in immediate
succession on 1971 August 26.  \citet{vandenber73lsand}
suggested that the variable might be either a supernova
or a flare star.  Although \citet{vandenber73lsand} did not
give the brightness of this object, it was estimated
to be 12.5 from their figure by \citet{rom77lsand}.

   \citet{sha73lsand} examined plates taken in the Crimean
Station of Sternberg Astronomical Institute and Latvian
Radio Astrophysical Observatory.  \citet{sha73lsand} succeeded
in obtaining one observation near the maximum and
the light curve of the fading part.  \citet{sha73lsand}
noted the presence of a star of 21--22~mag on Palomar Observatory
Sky Survey (POSS).  Based on the large amplitude exceeding
8~mag, rapid fading (0.2 mag~d$^{-1}$) in the early fading
part and the very slow (less than 0.001 mag~d$^{-1}$) fading rate
in the late fading part, \citet{sha73lsand} stated that
the star was unlikely a supernova or a flare star.
The light curve, however, did not resemble those of
typical novae or dwarf novae and \citet{sha73lsand} suggested
that it might be a very distant nova (i.e. intergalactic nova)
if it was indeed a nova.

   \citet{rom77lsand} examined Asiago plates and
presented a rough light curve of the outburst
(probably unaware of the work by \cite{sha73lsand}).
\citet{rom77lsand} indicated that the variable was at
the limit of visibility ($\sim$20.5~mag) on POSS
and that color was almost white.  \citet{rom77lsand} excluded
a flare star based on the light curve and also a supernova
based on the absence of a galaxy near the star.
\citet{rom77lsand} concluded that this object is probably
a dwarf nova of UV Per type.\footnote{
   UV Per was considered to be the prototype of dwarf novae
   with large-amplitude and rare outbursts at that time
   (cf. \cite{pet60DNe}).
   WZ Sge was considered as a recurrent nova and the concept
   of WZ Sge-type dwarf novae was not present.
   See \citet{kat15wzsge} for a modern review of WZ Sge-type
   dwarf novae.
}

   Following \citet{rom77lsand}, \citet{mei77lsand}
studied Sonneberg plates (probably also unaware of the work
by \cite{sha73lsand}) and constructed a light curve.
\citet{mei77lsand} concluded that the star was clearly
a fast nova and could not be a supernova due to the absence
of a galaxy near the star.  \citet{mei77lsand} excluded
a long-period dwarf nova (like UV Per) based on the facts:
(1) the amplitude was larger than 8~mag [\citet{mei77lsand}
even suggested that the object on POSS was a unrelated
one], (2) the decline after the maximum was too fast
and (3) no further outbursts were observed.
\citet{mei77lsand} suggested that this object was probably
a very bright nova in the halo of M~31.

   \citet{sha78lsand} and his colleagues examined materials
and found new records during the outburst close to the maximum
in the collection of Odessa Observatory.
Precise astrometry of the outbursting object using 
the materials at Latvian Radio Astrophysical Observatory
indicated the identity with the object on POSS.
Based on the large (9~mag) amplitude,
exceeding those of dwarf novae, \citet{sha78lsand} considered
that the object should be regarded as a fast nova despite
its small amplitude for a nova.  \citet{sha78lsand} also
remarked that the supposed nova did not follow
the maximum magnitude relation with decline time
for M~31 novae \citep{sha89m13nova}, and suggested that either
the relation was broken or the object was
an intergalactic nova 100--150~kpc from the Sun.
This classification by \citet{sha78lsand} was adopted
in \citet{due87novaatlas} and LS And was classified as a fast nova
in General catalogue of variable stars (GCVS: \cite{GCVS}).
In GCVS version 4.2 for extragalatic variables, LS And was
also given a name M31V0002 probably reflecting the possibility
of an object in M~31.

\begin{table*}
\caption{Observations of the 1971 outburst of LS And.}
\label{tab:obs1971}
\begin{center}
\begin{tabular}{ccc|ccc|ccc}
\hline\hline
JD\commenta & mag\commentb & source\commentc &
JD\commenta & mag\commentb & source\commentc &
JD\commenta & mag\commentb & source\commentc \\
\hline
179 & [19.0 & 3 & 223.497 & 18.30 & 2 & 292 & [19.0 & 3 \\
183.468 & [20.0 & 2 & 224.511 & 18.56 & 2 & 294 & [19.0 & 3 \\
183.508 & [13.6 & 5 & 225.541 & 18.30 & 2 & 296 & [19.0 & 3 \\
187.479 & 12.7: & 5 & 235 & 18.5 & 3 & 298 & [19.0 & 3 \\
187.508 & 11.7: & 5 & 236.248 & 18.80 & 2 & 300 & 19.0: & 3 \\
190 & 12.5 & 1 & 237.261 & 18.83 & 2 & 302 & 19.0 & 3 \\
191 & 13.8* & 4 & 238.405 & 18.83 & 2 & 304 & 19.0 & 3 \\
191.504 & 13.60 & 2 & 239.392 & 18.83 & 2 & 305.304 & 18.83 & 2 \\
193 & 14.0* & 4 & 240 & 18.7 & 3 & 308 & 19.0 & 3 \\
193.476 & 14.1: & 5 & 240.407 & 18.83 & 2 & 320 & 19.0: & 3 \\
193.507 & 14.1: & 5 & 242 & 18.7 & 3 & 324 & 19.0 & 3 \\
195.492 & 14.5:: & 5 & 245 & 18.5 & 3 & 332 & [19.0 & 3 \\
195.515 & 14.5:: & 5 & 245.339 & 19.0 & 2 & 335.238 & 18.8: & 2 \\
208 & 15.85 & 4 & 246.254 & 18.83 & 2 & 353.24 & 19: & 2 \\
209 & 14.9 & 3 & 249 & 18.5 & 3 & 570.392 & 19.2: & 2 \\
209.359 & 15.80 & 2 & 249.276 & [18.8 & 2 & 575.408 & 19.2: & 2 \\
210 & 16.25 & 4 & 252.434 & 18.8: & 2 & 655.286 & [18.3 & 2 \\
210.499 & 15.98 & 2 & 254.519 & [18.3 & 2 & 681.291 & 19.0 & 2 \\
212 & 16.4 & 3 & 263 & 18.5: & 3 & 682.168 & [19.2 & 2 \\
213.486 & 17.54 & 2 & 266.367 & 18.8 & 2 & 684.233 & 19.4 & 2 \\
214 & 17.6* & 4 & 268.427 & 18.83 & 2 & 685.201 & [19.2 & 2 \\
215 & 17.8* & 4 & 271 & 19.0 & 3 & 688.219 & 19.4 & 2 \\
217 & 18.0* & 4 & 276 & [19.0 & 3 & 983 & 20.0: & 5 \\
217.367 & 18.30 & 2 & 276.284 & 18.83 & 2 & 987 & 20.0: & 5 \\
220.358 & 18.33 & 2 & 277.396 & 18.8: & 2 & 2105 & 20: & 5 \\
221.545 & 18.38 & 2 & 278.308 & 18.9 & 2 & & & \\
222.4 & 18.43 & 2 & 280 & [19.0 & 3 & & & \\
\hline
\multicolumn{9}{l}{\commenta $\,$ JD$-$2441000.} \\
\multicolumn{9}{l}{\commentb $\,$ [ upper limits.  : uncertain.  * eye estimate from the published figure.} \\
\multicolumn{9}{l}{\commentc $\,$ 1: \citet{vandenber73lsand}, 2: \citet{sha73lsand}, 3: \citet{rom77lsand}, } \\
\multicolumn{9}{l}{\phantom{\commentc} 4: \citet{mei77lsand}, 5: \citet{sha78lsand}.} \\
\end{tabular}
\end{center}
\end{table*}

\begin{figure*}
\begin{center}
\includegraphics[width=14cm]{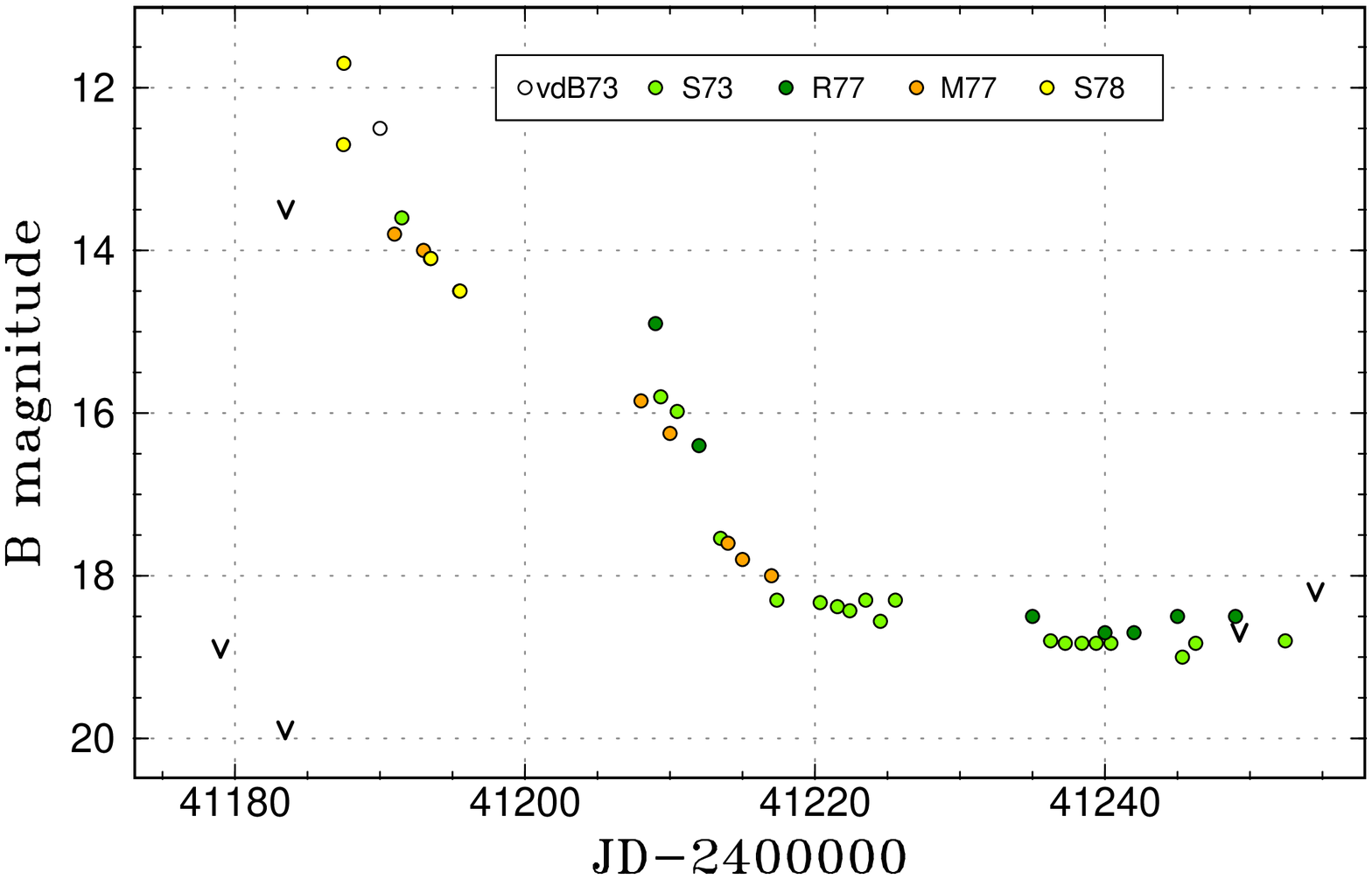}
\caption{
  Light curve of the 1971 outburst of LS And using the data
  in table \ref{tab:obs1971}.  The sources are
  vdB73 \citep{vandenber73lsand}, S73 \citep{sha73lsand},
  R77 \citep{rom77lsand}, M77 \citep{mei77lsand} and S78 \citep{sha78lsand}.
  The ``v'' symbols represent upper limits.
}
\label{fig:lsand1971lc}
\end{center}
\end{figure*}

   Although most professional astronomers considered or
treated LS And as a nova \citep{DownesCVatlas1,szk94oldnovaIR,
col09novapreoutburst,eva14novaWISE,ozd18novaMMRD},
and some suspected to be an X-ray nova \citep{ros99XNcandidate} or
a recurrent nova \citep{due88v394cra,pag14RNcand},
I may have been the first to become confident
that this should be a large-amplitude dwarf nova
after knowing this object in the freshly published work by
\citet{due87novaatlas}.  A part of the atmosphere in
the late 1980s among amateur astronomers was already told
in \citet{kat22stageA}.
Visual monitoring of LS And for a new outburst already started
in 1987 by VSOLJ members, and then by observers around the world.
Although results have not been fruitful for decades
[now exceeding 6000 observations without detecting
an outburst in the American Association of
Variable Stars (AAVSO)\footnote{
   $<$http://www.aavso.org/data-download$>$.
}; I myself had more than 200 non-detection
visual observations when I was an amateur astronomer],
I consistently considered LS And as a candidate
WZ Sge star \citep{kat01hvvir,kat02v592her}.
I expected that the Gaia satellite would clarify the nature of
LS And, but there was no parallax information in
Gaia DR2 \citep{GaiaDR2}.  The blue color (Gaia $B-R$=$+$0.25)
and a large proper motion were, however, sufficient to convince
me of the dwarf nova-type nature.  The parallax in
Gaia EDR3 \citep{GaiaEDR3} was not conclusive, probably due to
the faintness of this object.  The color in Gaia EDR3 was
even bluer ($B-R$=$-$0.06).

\begin{figure*}
\begin{center}
\includegraphics[width=14cm]{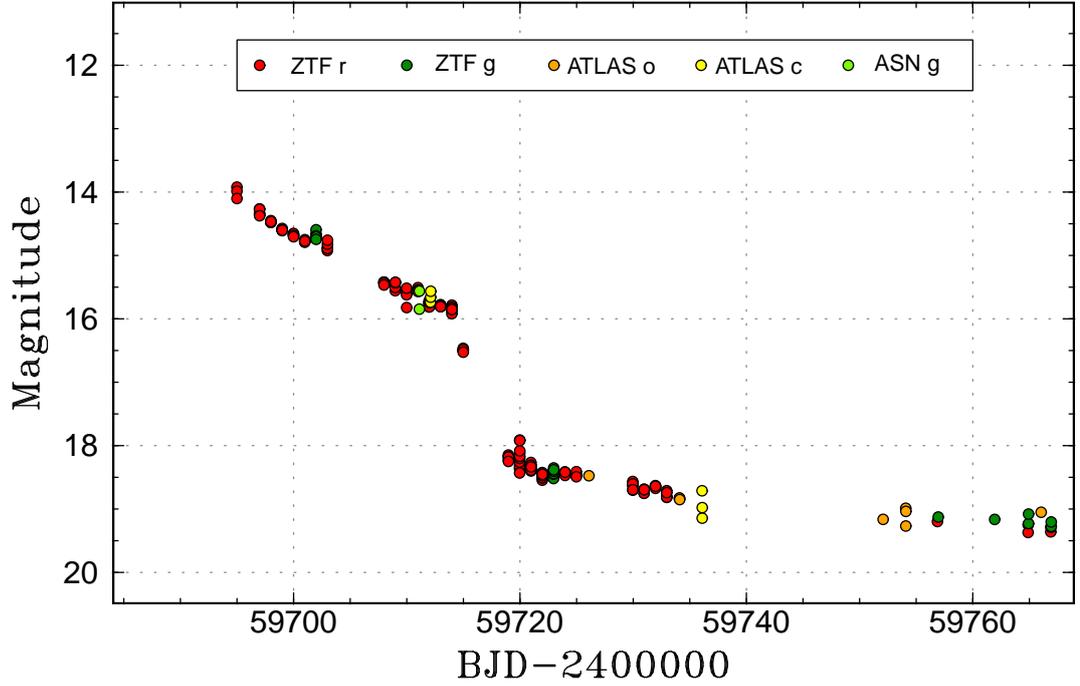}
\caption{
  Light curve of the 2022 outburst of LS And using ZTF, ATLAS
  and ASAS-SN data.  There were no upper limit observations
  before the initial detection.
}
\label{fig:lsand2022lc}
\end{center}
\end{figure*}

\begin{figure*}
\begin{center}
\includegraphics[width=14cm]{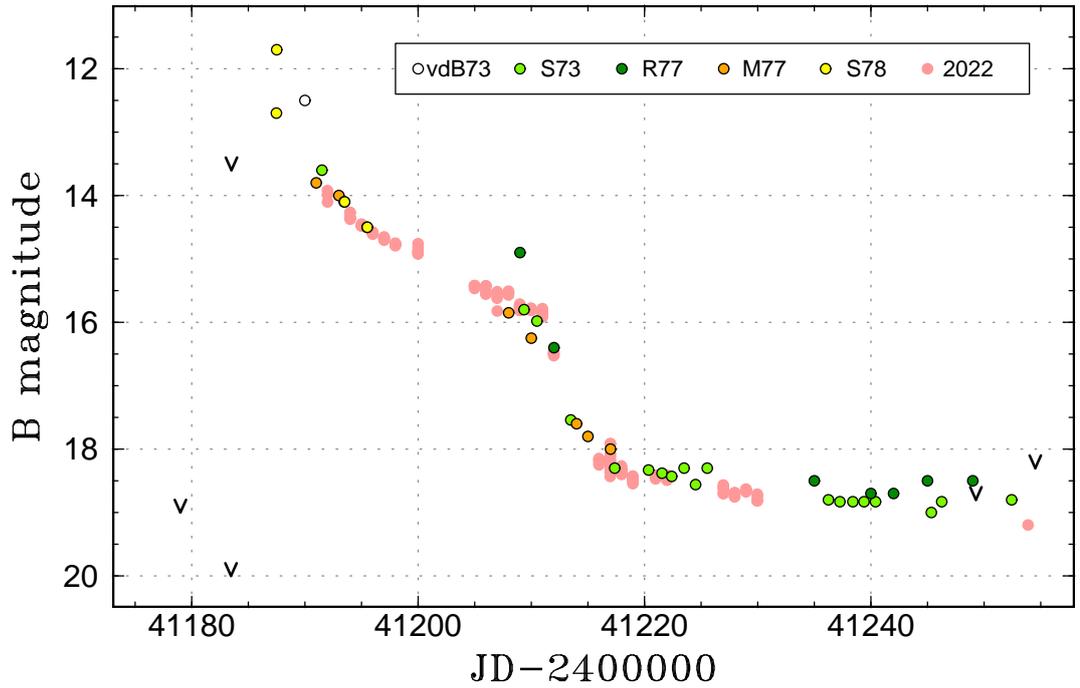}
\caption{
  Comparison of light curves of the 1971 and 2022 outburst of
  LS And.  The symbols for the 1971 observations are the same
  as in figure \ref{fig:lsand1971lc}.  The 2022 data
  (ZTF $r$ magnitudes) were shifted by 18503~d.
}
\label{fig:lsandcomp}
\end{center}
\end{figure*}

\begin{figure*}
\begin{center}
\includegraphics[width=14cm]{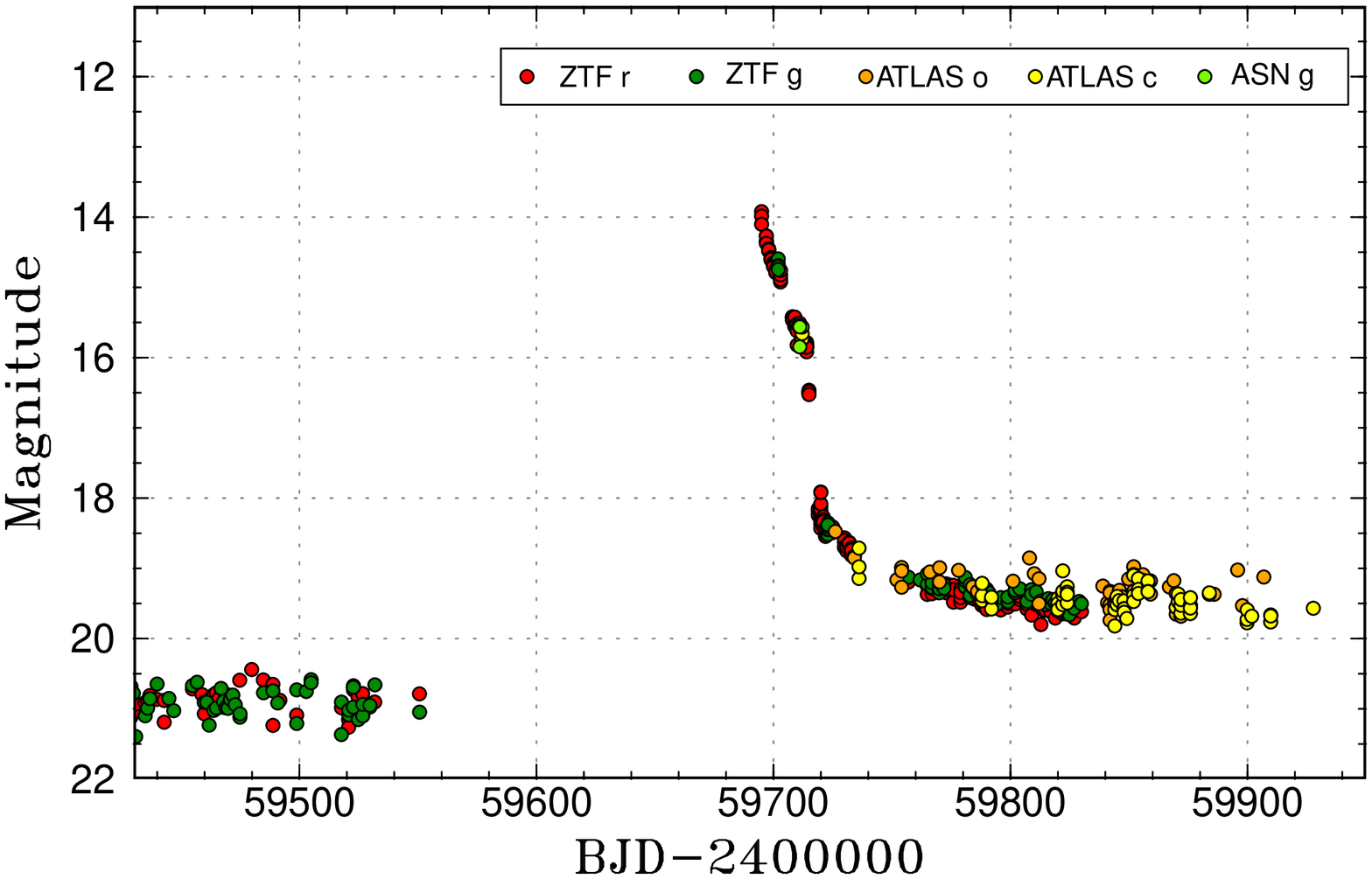}
\caption{
  Long-term light curve of the 2022 outburst of LS And.
  The symbols are the same as in figure \ref{fig:lsand2022lc}.
}
\label{fig:lsand2022long}
\end{center}
\end{figure*}

   The ``moment'' arrived like lightening when I was examining
light curves obtained by
the Zwicky Transient Facility (ZTF: \cite{ZTF})\footnote{
   The ZTF data can be obtained from IRSA
$<$https://irsa.ipac.caltech.edu/Missions/ztf.html$>$
using the interface
$<$https://irsa.ipac.caltech.edu/docs/program\_interface/ztf\_api.html$>$
or using a wrapper of the above IRSA API
$<$https://github.com/MickaelRigault/ztfquery$>$.
}.  It was when I started examining light curves of
recent ZTF data.  As usual, I was looking at the table
of dwarf novae listed in alphabetical order, and almost
unconsciously typed LS~And (as a matter of fact, I already did
not pay special attention to this object regularly since
I knew that it had been well monitored by amateur observers
and considered that no missed outburst would be expected in
the ZTF data).  The reason why I specially selected LS And
was unknown, but the light curve on the display was a familiar
one of a WZ Sge star.  I initially considered that I entered
a name of a different well-known WZ Sge star (almost unconsciously
as a routine work), but realized that it was ``LS And''.
Unthinkable!  I initially could not believe my eyes,
but it was indeed LS And and I almost automatically issued
vsnet-alert 27267\footnote{
  $<$http://ooruri.kusastro.kyoto-u.ac.jp/mailarchive/vsnet-alert/27267$>$.
}, even without sufficient patience for waiting the result of
a query to the All-Sky Automated Survey for Supernovae
(ASAS-SN) Sky Patrol data \citep{ASASSN,koc17ASASSNLC}.
My emotion at that time may have been similar to
a situation when I encountered a rare bird which I could not
believe (cf. \cite{kat22stageA}).  Birders will agree.

   In the world of birders, it must have become the busiest
moment after any discovery --- one needs to locate the bird
and take images or recordings sufficient for a proof
of the existence of a rare bird.  The case for the detection
of the 2022 outburst of LS And was different.  There was no
special care for preserving the data shown on the display,
and I went to the library (fortunately very close) to search
the light curve of the 1971 outburst, which still stayed deep
in my memory even after decades.

   So it's time to return to science.
In table \ref{tab:obs1971}, I summarized photometric data
for the 1971 outburst.  The magnitudes were all photographic
(equivalent to $B$).  Magnitudes with * were estimated by
my eyes from the figure in \citet{mei77lsand}, which are probably
correct to $\pm$1~d and $\pm$0.1~mag.  The magnitude for
JD$=$190 was similarly estimated from a published figure
by \citet{rom77lsand}.  \citet{mei77lsand} claimed that
the object was estimated too bright by \citet{rom77lsand}.
The light curve drawn from these data is presented in
figure \ref{fig:lsand1971lc}.  This is not much different from
the one published in \citet{sha78lsand}, but is worth presenting
here since \citet{sha78lsand} is difficult to reach.

   The 2022 light curve is shown in figure \ref{fig:lsand2022lc}.
It is very clear that the 1971 and 2022 light curves
are very similar: plateau-type fading lasting for $\sim$20~d
followed by rapid decline and subsequent slow fading.
They are typical WZ Sge-type outbursts without rebrightening
(type D superoutburst in \cite{kat15wzsge}).
It is also well-known that the same WZ Sge star tends to
repeat the same type of rebrightening \citep{kat15wzsge}
and LS And is also the case.  Although the mechanism of
rebrightening(s) is not yet well understood,
empirical relationship shows that WZ~Sge stars without
rebrightening are mostly objects near the period minimum
of cataclysmic variables, but before reaching it
(figure 17 in \cite{kat15wzsge}).
The orbital period of LS And is thus expected to be
within 0.053--0.060~d.  The fading rate of the plateau phase
(BJD 2459696--2459714.5) was 0.089(1) mag~d$^{-1}$,
which corresponds to $\log t_{\rm d}$=1.05, a typical
value for a WZ Sge star without rebrightening and not
resembling a period bouncer (see figure 87 in \cite{Pdot5}).
A comparison between the 1971 and 2022 outbursts
is shown in figure \ref{fig:lsandcomp} (from now on,
I treat all photometric bands in visual wavelengths
almost identical with $V$, which is a good approximation
for a WZ Sge star in outburst).
These outbursts were almost exactly
the same and the interval of these two outburst
was 18503~d (=50.66~yr).
This comparison suggests that the 2022 outburst would
not have started before JD 2459682 (2022 April 12).
Definitely a sigh! (particularly for amateur observers)
considering the almost no evening visibility of this object
in mid-April.

   People may wonder if these outburst could be those of
an SU UMa star rather than a WZ Sge star, and how I can be
confident about the classification without observation
of early superhumps (cf. \cite{kat15wzsge}).
I show a long-term light curve of the 2022 outburst in
figure \ref{fig:lsand2022lc}.  The object was brighter
by 1.5~mag after the outburst.  The post-outburst phenomenon
is a long fading tail, which is characteristic to
a WZ Sge-type outburst and not seen in an SU UMa star.
The presence of the same phenomenon was also reported
after the 1971 outburst \citep{sha73lsand}.\footnote{
   It might be interesting to leave a remark that the figure
   in \citet{sha73lsand} dealt with this phenomenon
   rather than the shape of the outburst.
   Please have a look at his figure if you have a chance
   too see this reference.
}
Before the outburst plateau, there was a phase with
more rapid fading (more evident in the 1971 light curve
and only one day in the 2022 one).  This feature is
commonly seen in WZ Sge-type outbursts and is referred
to as a viscous decay phase.  Early superhumps are expected
during this phase if the binary has a sufficient
inclination \citep{kat15wzsge,kat22wzsgeMV}.

   The peak magnitude probably requires re-examination.
Although most literature gives 11.7~mag as the maximum
for LS And, it is evident from table \ref{tab:obs1971}
that this magnitude was uncertain (``:'' usually means that
the object is close to the limit of photographic materials
or the quality of the photograph is poor)
and was the brighter one
of two uncertain observations (11.7 and 12.7~mag) 
only 40~min apart.  It looks more likely that the true brightest
observation was close to their average (12.2~mag).
The outburst amplitude based on this value is 8.8~mag
using the ZTF data before the 2022 outburst.
The true peak would have been brighter,
though, since there was a 4~d observational gap before
the first observation of the outburst (but see the discussion
below).

   As seen from the 2022 observations, the magnitude when
ordinary superhumps should appear following
the viscous decay phase was 14.3~mag.
In ordinary WZ Sge stars, the absolute magnitude ($M_V$) when
ordinary superhumps appear is $+$5.4
(for an average inclination of 1~radian) \citep{kat22wzsgeMV}.
Using this value as the standard candle, the distance modulus
of LS And is estimated to be 8.9.  The observed peak (12.2~mag)
in 1971 corresponds to $M_V$=$+$3.3.  The quiescent magnitude
(21.0~mag, ZTF data) corresponds to $M_V$=$+$12.1.
The difference (6.7~mag) between quiescent magnitude and
the magnitude when ordinary superhumps appear is typical
for a (non-period bouncer) WZ Sge star
(see fig. 23 in \cite{kat15wzsge}; \cite{tam20v3101cyg}).
Other properties of LS And are expected to be similar to
those of typical WZ Sge stars.

   The detection of the 2022 outburst of LS And brought
a some kind of despair to observers who had been expecting
to see a fresh outburst for decades.
Could there be a possibility that LS And silently underwent
outbursts more frequently only around solar conjunctions?
This was indeed the case of the SU UMa star VY~Aqr located
close to the ecliptic.
Despite the mean interval of superoutbursts of
less than 2~yr, this object was not recorded in superoutburst
between 1994 and 2006, and between 2008 and 2020.
It was most likely that superoutbursts in this object occurred
around solar conjunctions and were not recorded.
Although similar things may have happened in LS And
at least in the past, modern deep observations
such as ZTF should have detected the object during
a fading tail if there was a missed superoutburst.
There was no indication of such a detection in the ZTF data
since 2018, and the outburst interval should be longer
than 5~yr.  The fading tail lasted more than a year
\citep{sha73lsand}.  \citet{sha78lsand} described that
the object returned to practically the same level
before the outburst after 5.5~yr, although this description
may have assumed a nova-type light curve and could have
overestimated the duration of the fading tail.
Considering these values and considering that parameters
of LS And are similar to those of typical WZ Sge stars,
the next major outburst would be expected after a decade
or even more [see figure 5 in \citet{kat15wzsge} for
the distribution of outburst intervals in WZ Sge stars].
By comparing the recorded peak $M_V$=$+$3.3 (in 1971)
with the statistics of known WZ Sge stars
(figure 10 in \cite{tam20v3101cyg}), it appears that
the true peak in 1971 was not missed after a considerable
delay (i.e. the object was unlikely to have reached 11.0~mag
even at the true peak).  The next superoutburst would
also be around 12.2~mag.  There are, however, exceptional
objects like V3101 Cyg \citep{tam20v3101cyg,ham21DNrebv3101cyg}
and there may be an unexpected phenomenon even after
the outburst.
In the post-outburst data of LS And, 0.5~mag brightening
lasting for 10--20~d and starting around JD 2459852
was present (figure \ref{fig:lsandcomp}).
This might suggest that LS And
is still active in the post-superoutburst phase and
would be worth observing before it finally returns to
quiescence.

\section*{Acknowledgements}

This work was supported by JSPS KAKENHI Grant Number 21K03616.
The author is grateful to the ZTF, ATLAS and ASAS-SN teams
for making their data available to the public.
I am grateful to VSOLJ, AAVSO and VSNET observers for
reporting observations and to Naoto Kojiguchi for
helping downloading the ZTF data.

Based on observations obtained with the Samuel Oschin 48-inch
Telescope at the Palomar Observatory as part of
the Zwicky Transient Facility project. ZTF is supported by
the National Science Foundation under Grant No. AST-1440341
and a collaboration including Caltech, IPAC, 
the Weizmann Institute for Science, the Oskar Klein Center
at Stockholm University, the University of Maryland,
the University of Washington, Deutsches Elektronen-Synchrotron
and Humboldt University, Los Alamos National Laboratories, 
the TANGO Consortium of Taiwan, the University of 
Wisconsin at Milwaukee, and Lawrence Berkeley National Laboratories.
Operations are conducted by COO, IPAC, and UW.

The ztfquery code was funded by the European Research Council
(ERC) under the European Union's Horizon 2020 research and 
innovation programme (grant agreement n$^{\circ}$759194
-- USNAC, PI: Rigault).

This work has made use of data from the Asteroid Terrestrial-impact
Last Alert System (ATLAS) project. The Asteroid Terrestrial-impact
Last Alert System (ATLAS) project is primarily funded to search for
near earth asteroids through NASA grants NN12AR55G, 80NSSC18K0284,
and 80NSSC18K1575; byproducts of the NEO search include images and
catalogs from the survey area. This work was partially funded by
Kepler/K2 grant J1944/80NSSC19K0112 and HST GO-15889, and STFC
grants ST/T000198/1 and ST/S006109/1. The ATLAS science products
have been made possible through the contributions of the University
of Hawaii Institute for Astronomy, the Queen's University Belfast, 
the Space Telescope Science Institute, the South African Astronomical
Observatory, and The Millennium Institute of Astrophysics (MAS), Chile.

\section*{List of objects in this paper}
\xxinput{objlist.inc}

\section*{References}

We provide two forms of the references section (for ADS
and as published) so that the references can be easily
incorporated into ADS.

\renewcommand\refname{\textbf{References (for ADS)}}

\newcommand{\noop}[1]{}\newcommand{\hyphalt}{-}

\xxinput{lsandaph.bbl}

\renewcommand\refname{\textbf{References (as published)}}

\xxinput{lsand.bbl.vsolj}

\end{document}